\documentclass[english,twocolumn]{revtex4}
\usepackage[T1]{fontenc}
\usepackage[latin1]{inputenc}
\usepackage{amsmath}
\usepackage{graphicx}
\usepackage{amssymb}

\makeatletter

\renewcommand{\vec}[1]{\mathbf{#1}}

\usepackage{babel}
\makeatother
\begin{document}

\newcommand{\bIntrinsicVec}{\vec{b}^{\left(\mathrm{i}\right)}}

\newcommand{\scatteringAngle}{\gamma}

\newcommand{\gammaPlusTildeN}{\tilde{\gamma}_{\left(+\right)}}

\newcommand{\gammaMinusTildeN}{\tilde{\gamma}_{\left(-\right)}}

\newcommand{\gammaPlusMinusTildeN}{\tilde{\gamma}_{\left(\pm\right)}}

\newcommand{\gammaHomTildeN}{\tilde{\gamma}_{0}}

\newcommand{\nTwoD}{n_{2\mathrm{D}}}

\newcommand{\polDensity}{\vec{s}^{E}}

\newcommand{\polDensityElem}{s^{E}}

\newcommand{\polDensityElemDot}{\dot{s}^{E}}

\newcommand{\polDensityIso}{\vec{s}^{\lblIso}}

\newcommand{\polDensityAniso}{\vec{s}^{\lblAniso}}

\newcommand{\polDensityElemAniso}{s^{\lblAniso}}

\newcommand{\polDensityElemDotAniso}{\dot{s}^{\lblAniso}}

\newcommand{\polDensityElemEq}{s^{\mathrm{eq}}}

\newcommand{\polDensityElemDotEq}{\dot{s}^{\mathrm{eq}}}

\newcommand{\polDensityTot}{\vec{s}}

\newcommand{\polDensityElemTot}{s}

\newcommand{\polDensityElemDotTot}{\dot{s}}

\newcommand{\polDensityDotTot}{\vec{\dot{s}}}

\newcommand{\tauSpin}{\tau_{\mathrm{s}}}

\newcommand{\tauSpinApp}{\tilde{\tau}_{\mathrm{s}}}

\newcommand{\tauSpinInPlane}{\tau_{xy}}

\newcommand{\tauSpinz}{\tau_{z}}

\newcommand{\tauSpinGeomMean}{\tauSpin}

\newcommand{\invTauSpinTensor}{{\overleftrightarrow{\tau}\hspace{-.5mm}{}_{\mathrm{s}}^{-1}}}

\newcommand{\azimuthalAngle}{\varphi}

\newcommand{\unitStep}{\Theta}

\newcommand{\DOS}{\nu}

\newcommand{\ZeemanX}{B}

\newcommand{\fSpinEisoSol}{\fSpinE_{\lblIso}}

\newcommand{\fSpinEanisoCorr}{\fSpinE_{\lblAniso}}

\newcommand{\anisoPrefix}{\delta}

\newcommand{\lblIso}{\mathrm{is}}

\newcommand{\lblAniso}{\mathrm{ai}}

\newcommand{\jSpin}[2]{j_{#2}^{#1}}

\newcommand{\jChargeComp}[1]{j_{#1}^{\mathrm{c}}}

\newcommand{\jSpLblPhimu}{(1)}

\newcommand{\jSpLblPhiC}{(2)}

\newcommand{\Eparam}{\beta}

\newcommand{\FTarrowDisplayed}{\stackrel{\mathrm{FT}}{\longrightarrow}}

\newcommand{\Ls}{L_{\mathrm{s}}}

\newcommand{\Lsz}{\Ls^{z}}

\newcommand{\ZeemanHanle}{\omega_{\mathrm{L}}}

\newcommand{\ZeemanHanleHWHM}{\ZeemanHanle^{\mathrm{HWHM}}}

\newcommand{\tauSpinxyzFac}{\alpha}

\newcommand{\HWHMdimAnFunc}{g}

\newcommand{\distSpinDiff}{d}

\newcommand{\sqrtDiscriminant}{T}

\newcommand{\invBDdecayLength}{q_{\mathrm{s}}}

\newcommand{\bsigma}{\boldsymbol{\sigma}}

\newcommand{\vDrift}{\vec{v}_{\mathrm{dr}}}

\newcommand{\vDriftAbs}{v_{\mathrm{dr}}}

\newcommand{\brelaxNoDim}{\beta}

\newcommand{\qDr}{q}

\newcommand{\BNodim}{b}

\newcommand{\sqrtsqrtB}{p}
 
\newcommand{\qRe}{\kappa_{R}}

\newcommand{\qIm}{\kappa_{I}}

\newcommand{\driftField}{\vec{b}_{\mathrm{dr}}}

\newcommand{\curvHanle}{c}

\title{Hanle Effect near Boundaries}

\author{Hans-Andreas Engel}

\affiliation{Department of Physics, Harvard University, Cambridge, Massachusetts
02138 }

\begin{abstract}
The Hanle effect describes suppression of spin polarization due to
precession in a magnetic field. This is a standard spintronics tool
and it gives access to the spin lifetime of samples in which spins
are generated homogeneously. We examine the Hanle effect when spins
are generated at a boundary of a diffusive sample by the extrinsic
spin Hall effect. We show that the Hanle curve is spatially dependent
and that the {}``apparent'' spin lifetime, given by its inverse
half-width, is shorter near the boundary even if the spin relaxation
rate is homogenous. 
\end{abstract}
\maketitle
The goal of spintronics is to generate and manipulate spin populations
on time scales limited by the spin lifetime. One can access the spin
population optically, since selection rules allow optical pumping
and detection of spins in materials \cite{OpticalOrientation}; interesting
alternatives are magnetic materials or materials with spin-orbit interaction,
providing access to spins with standard microelectronic devices \cite{Wolf_Spintronics,spintronicsBook}.
To characterize a given sample, it is essential to determine its spin
lifetime $\tauSpin$, which depends on the microscopic properties
of the sample. One can determine $\tauSpin$ of a homogeneous sample
using the Hanle effect \cite{OpticalOrientation} as follows, even
if time-resolved measurements are not available. If there is no spin
precession, a spin polarization simply decays with time $\tauSpin$.
However, if a magnetic field $B$ perpendicular to the spin polarization
axis is applied, there is a competing relaxation mechanism: spins
will precess in that magnetic field with Larmor frequency $\ZeemanHanle\propto B$.
If the magnetic field is sufficiently large, such that the spin can
precess many times within its lifetime, this will randomize the spin
and suppress the spin polarization. This competing spin relaxation
mechanism becomes effective for $\ZeemanHanle\gtrsim1/\tauSpin$---thus
$\tauSpin$ can be extracted by measuring the inverse width of the
so-called Hanle curve $s_{z}(\ZeemanHanle)$.

In recent experiments by Kato \emph{et al.}~\cite{KatoSpinHall},
a spatially dependent spin polarization $s_{z}$ was induced via the
extrinsic spin Hall effect~\cite{DPpolarization,Hirsch99,EHR_extrinsic,TseDasSarmaExtrinsic}
and measured via Kerr microscopy. The width of the Hanle curves $s_{z}(\ZeemanHanle,\,\vec{r})$
was described with a spatially-dependent spin lifetime $\tauSpinApp(\vec{r})$.
Rather strikingly, it was found that $\tauSpinApp$ is several times
smaller near the sample edge than $10\:\mathrm{\mu m}$ away from
the edge. In this article we calculate the Hanle curves and show that
such a suppression of $\tauSpinApp$ near the edge can result from
spin diffusion, even if the spin relaxation rate $\tauSpin^{-1}$
is spatially \emph{homogeneous}. 

\begin{figure}
\begin{center}\includegraphics[%
  width=85mm,
  keepaspectratio]{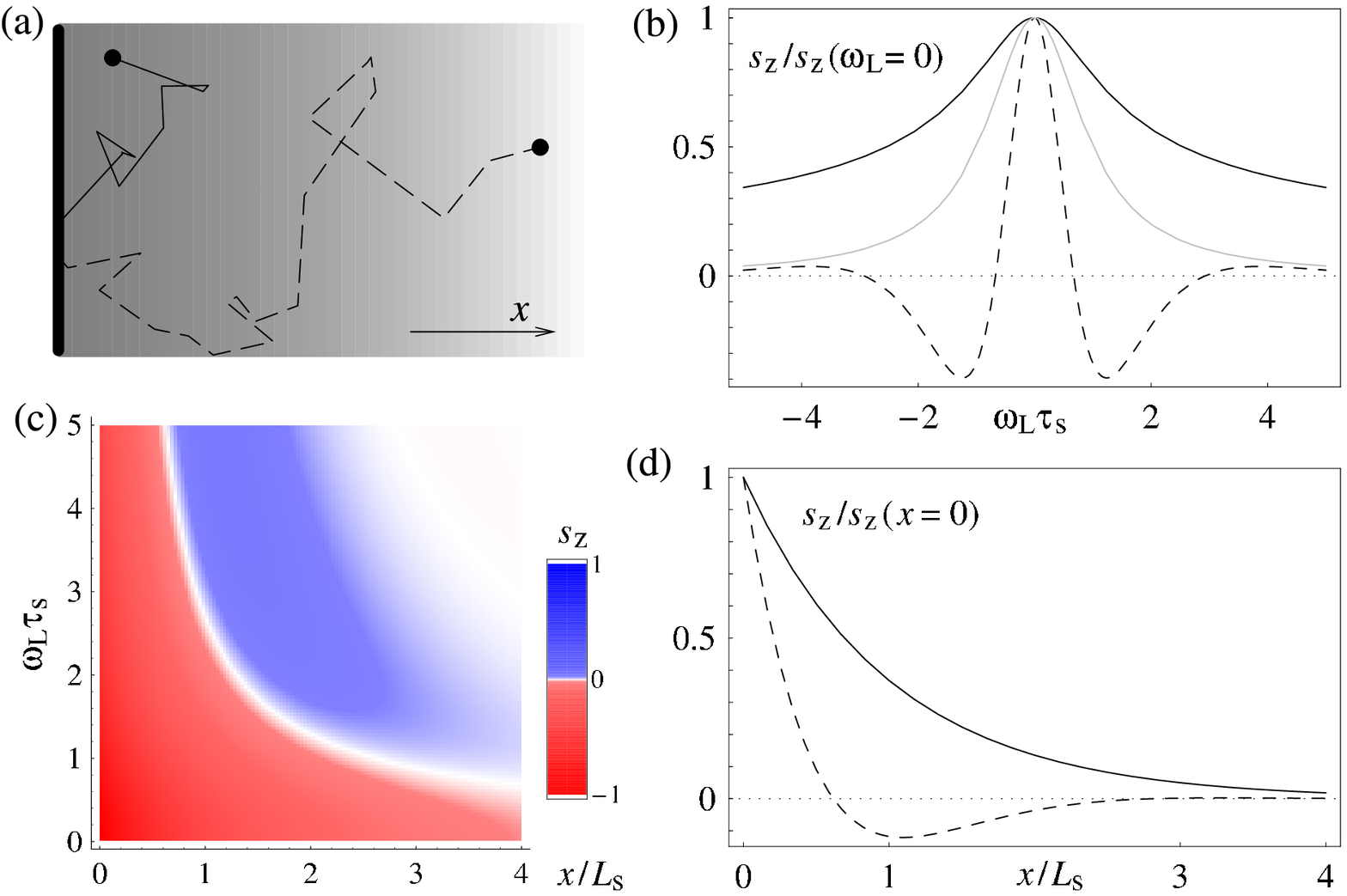}\vspace{-4mm}${}$\end{center}

\caption{\label{cap:hanle}Spin density $s_{z}(x,\,\ZeemanHanle)$ near a
boundary of a diffusive system, given by Eq.~(\ref{eq:szofxGen})
for $\tauSpinGeomMean=\tauSpinInPlane=\tauSpinz$ and $\invBDdecayLength=0$.
(a) When spins are generated at the boundary and then diffuse into
the sample, spins closer to the boundary had on average less time
to precess in the magnetic field $\vec{B}=B\vec{\hat{x}}$ than those
further away. (b) Therefore, the Hanle curve $s_{z}(\ZeemanHanle\propto B)$
close to the boundary ($x=0$, solid line) becomes broader, while
away from the boundary ($x=4\Ls$, dashed line) it becomes narrower
than in the case of homogeneous spin generation (gray line). (c) Sufficiently
far from the boundary the Hanle curve develops {}``sidelobes'' \cite{JohnsonSilsbeeSidelobe,JohnsonSilsbeeBloch}
where the spin polarization changes sign; here $s_{z}$ is plotted
in units of $j_{x}^{z}\sqrt{\tauSpinGeomMean/D}$. (d) Spins generated
at the boundary diffuse into the sample in the absence of a magnetic
field and eventually become suppressed due to spin relaxation ($\ZeemanHanle=0$,
solid line), while in a magnetic field ($\ZeemanHanle=5\tau_{\mathrm{s}}^{-1}$,
dashed line) spin precession further suppresses spin polarization.}
\end{figure}

The physical picture for this spatial dependence of $\tauSpinApp$
is as follows {[}see Fig.~\ref{cap:hanle}(a),(b){]}. Spins are generated
at the boundary and then diffuse into the bulk of the sample. In a
magnetic field, the spins observed at a small distance $x$ were (on
average) generated a short time ago and did not yet precess much in
the magnetic field. Therefore, they have a larger $s_{z}$ than one
would expect for the homogeneous case with a bulk generation mechanism
(e.g., optical pumping). This means that the linewidth as function
of $B$ is larger and the spin lifetime seems smaller. Conversely,
the spins observed far from the boundary, required a rather long time
to get there and were able to precess longer in the magnetic field.
Therefore, the value of $s_{z}$ is more strongly suppressed by $B$,
the linewidth becomes narrower, and the spin lifetime appears longer.

A similar situation is found when the dominating spin transport mechanism
is the drift induced by charge currents \cite{Crooker05,KatoL,ShiInhomogeneous,FinklerSHE}.
During the drift from the injection to the detection point over a
distance $r$, spins precess during time $t=r/\vDriftAbs$, where
$\vDriftAbs$ is the drift velocity. Because the precession angle
$\ZeemanHanle t$ is the same for each spin (neglecting diffusion),
multiple oscillations of $s_{z}$ were observed as function of $\ZeemanHanle$
\cite{Crooker05} or $r$ \cite{KatoL}.

To quantitatively describe the suppression of the apparent spin lifetime
$\tauSpinApp$ in a diffusive system, we now analyze the Hanle curves
for such systems. For this, we consider a magnetic field $\vec{B}=B\,\vec{\hat{x}}$,
which induces spin precession of electrons with Larmor frequency $\ZeemanHanle=g^{*}\mu_{\mathrm{B}}B/\hbar$,
with effective $g$-factor $g^{*}$ and Bohr magneton $\mu_{\mathrm{B}}$,
corresponding to a Zeeman coupling $H_{\mathrm{Z}}=\frac{1}{2}\, g^{*}\mu_{\mathrm{B}}\vec{B}\cdot\bsigma$.
We assume a sufficiently small magnetic field that orbital effects
are not important and that $\tauSpin^{-1}$ is independent of $B$.
The equation of motion for the spin density $\vec{s}$, including
spin precession, diffusion, and relaxation is  \begin{align}
\vec{\dot{s}} & =(g^{*}\mu_{\mathrm{B}}/\hbar)\,\vec{B}\times\vec{s}+D\Delta\vec{s}-\invTauSpinTensor\vec{s},\label{eq:Precession}\end{align}
 with a spatially independent spin diffusion constant $D$ and a
diagonal spin relaxation tensor $\invTauSpinTensor$ with components
$\{\tauSpinInPlane^{-1},\,\tauSpinInPlane^{-1},\,\tauSpinz^{-1}\}$;
note that the polarization $s_{x}$ decouples and so its spin relaxation
rate is actually not important here. Also, we define the geometrical
mean of the spin relaxation times as $\tauSpinGeomMean=\sqrt{\tauSpinInPlane\tauSpinz}$
and the spin diffusion length is $L_{\mathrm{s}}=\sqrt{D\tauSpinGeomMean}$.
We set $\tauSpinInPlane=\tauSpinxyzFac\tauSpinGeomMean$ and $\tauSpinz=\tauSpinxyzFac^{-1}\tauSpinGeomMean$
with some dimensionless constant $\tauSpinxyzFac$, e.g., $\tauSpinxyzFac=\sqrt{2}$
for Dyakonov-Perel spin relaxation and Rashba coupling \cite{DPrelax72,DPtime2D,Burkov04,LauFlatte_T1T2}. 

Next we assume that spin polarization is generated at a boundary plane.
This is the case for the extrinsic spin Hall effect \cite{KatoSpinHall,DPpolarization,Hirsch99,EHR_extrinsic,TseDasSarmaExtrinsic},
where an electrical current induces spin currents which in turn produce
spin polarization near sample edges due to extrinsic spin-orbit interaction.
We take a semi-infinite two- or three-dimensional system with $x\geq0$,
and an electric field $E_{y}$ applied along the $y$-direction. The
transverse spin current is $j_{x}^{z}=\sigma_{\mathrm{SH}}E_{y}$,
with spin Hall conductivity $\sigma_{\mathrm{SH}}$; microscopically
the spin current relaxes on the short transport lifetime $\tau\ll1/\ZeemanHanle$,
thus $\sigma_{\mathrm{SH}}$ does not depend on the weak magnetic
field. If spin is conserved at the boundary, there is no spin current
perpendicular to the boundary and the spin Hall current is compensated
by spin diffusion, i.e., $j_{x}^{z}=D\,\frac{\partial}{\partial x}\, s_{z}$
at $x=0$. More generally, we consider the boundary condition \begin{align}
\frac{\partial}{\partial x}\, s_{z}=\frac{j_{x}^{z}}{D}+\invBDdecayLength s_{z} & ,\qquad\frac{\partial}{\partial x}\, s_{y}=\invBDdecayLength s_{y},\label{eq:Boundary}\end{align}
which allows for spin relaxation at the edge, characterized by $\invBDdecayLength$,
and where we have taken $j_{x}^{y}=0$.

For other systems, where spins are generated at a boundary and then
precess in a field, Eqs.~(\ref{eq:Precession}), (\ref{eq:Boundary})
also apply and these systems show the same Hanle curves. D'yakonov
and Perel' \cite{DP_HanleSurface} considered the situation where
electron spins were optically generated using circularly polarized
light in a surface layer thinner than $\Ls$. Assuming that recombination
only takes place in this surface layer, it is taken into account via
$\invBDdecayLength$. Further, the degree of circular polarization
of the recombination radiation is proportional to $s_{z}$ at $x=0$,
so only the Hanle curve at the surface is experimentally accessible.
Such measurements were reported by Vekua \emph{et al}. \cite{Vekua_HanleSurfExp}.
Furthermore, Johnson and Silsbee~\cite{JohnsonSilsbeeBloch}, analyzed
a system where spins are injected from a ferromagnet into a paramagnet
at $x=0$. A second ferromagnet at a distance $x$ is used as a detector,
whose voltage is proportional to the spin polarization $s_{z}(x)$.
Fabrication of devices with different detector spacings then provides
electrical access to the spatially-dependent Hanle curve.

We now analyze the spin polarization in the stationary case $\vec{\dot{s}}=0$
by assuming that the spin relaxation rate $\invTauSpinTensor$ is
\emph{spatially independent}. With the ansatz $\vec{s}=\vec{s}_{0}e^{qx}$
we find the solutions of Eq.~(\ref{eq:Precession}) satisfying $\mathrm{Re}\, q<0$,
\begin{align}
q_{0,1} & =-\sqrt{\frac{\tauSpinInPlane+\tauSpinz\pm\sqrtDiscriminant}{2D\tauSpinInPlane\tauSpinz}},\\
\sqrtDiscriminant & =\sqrt{(\tauSpinInPlane-\tauSpinz)^{2}-4\tauSpinInPlane^{2}\tauSpinz^{2}\ZeemanHanle^{2}}.\end{align}
From the boundary conditions (\ref{eq:Boundary}), we obtain the position-dependent
Hanle curves\begin{align}
s_{y}(x,\, B) & =j_{x}^{z}\,\sum_{i=0,1}e^{q_{i}x}\,\frac{\left(-1\right)^{i}\tauSpinInPlane\tauSpinz\ZeemanHanle}{D\sqrtDiscriminant\,\left(q_{i}-\invBDdecayLength\right)},\label{eq:syofxGen}\\
s_{z}(x,\, B) & =j_{x}^{z}\,\sum_{i=0,1}e^{q_{i}x}\,\frac{\sqrtDiscriminant+\left(-1\right)^{i}(\tauSpinInPlane-\tauSpinz)}{2D\sqrtDiscriminant\,\left(q_{i}-\invBDdecayLength\right)}.\label{eq:szofxGen}\end{align}

For $\tauSpinInPlane=\tauSpinz=\tauSpinGeomMean$, Eq.~(\ref{eq:szofxGen})
simplifies considerably; using $q_{0,1}\Ls=-\sqrt{1\pm i\ZeemanHanle\tauSpinGeomMean}=-(\qRe\pm i\qIm)$
with $\qRe=\big(1+\sqrt{1+\ZeemanHanle^{2}\tauSpinGeomMean^{2}}\big)^{1/2}/\sqrt{2}$
and $\qIm=\ZeemanHanle\tauSpinGeomMean/2\qRe$, we find \begin{align}
s_{z}(x) & =-\frac{j_{x}^{z}\,\sqrt{\tauSpin}\,\big[\left(\qRe+\brelaxNoDim\right)\cos\qIm\distSpinDiff+\qIm\sin\qIm\distSpinDiff\big]}{\sqrt{D}\big[\left(\qRe+\brelaxNoDim\right)^{2}+\qIm^{2}\big]}\, e^{-\qRe\distSpinDiff},\label{eq:szofxAlpha1}\end{align}
where we have defined $\brelaxNoDim=\invBDdecayLength\Ls$. In the
special case of $x=0$ and $\tauSpinInPlane=\tauSpinz$, Eq.~(\ref{eq:szofxAlpha1})
agrees with the expression found when studying Hanle effect on surfaces~\cite{DP_HanleSurface,Vekua_HanleSurfExp};
while for $\invBDdecayLength=0$ and $\tauSpinInPlane=\tauSpinz$
it agrees with the result from Ref.~\onlinecite{JohnsonSilsbeeBloch}. 

Further, in the absence of the magnetic field, Eq.~(\ref{eq:szofxGen})
simplifies to $s_{z}=-j_{x}^{z}\sqrt{\tauSpinGeomMean/D}\, e^{-\sqrt{\alpha}x/\Ls}/(\sqrt{\alpha}+\brelaxNoDim)$.
Finally, for $\invBDdecayLength=0$, the integrated spin density corresponds
to the Hanle curve of a homogeneous system, \begin{equation}
\int dx\: s_{z}(x)=-\frac{j_{x}^{z}\,\tauSpinz}{1+\tauSpinGeomMean^{2}\ZeemanHanle^{2}}.\label{eq:szIntegrated}\end{equation}

In the experiments of Ref.~\onlinecite{KatoSpinHall}, the apparent
spin lifetime $\tauSpinApp(x)$ is extracted by assuming a Lorentzian
Hanle curve {[}Eq.~(\ref{eq:szIntegrated}){]} at each position,
then $\tauSpinGeomMean$ can be found as the half-width at half-maximum
of $s_{z}(\ZeemanHanle)$. Correspondingly, we now take Eq.~(\ref{eq:szofxGen})
and solve $\frac{1}{2}s_{z}(x,\,\ZeemanHanle=0)=s_{z}(x,\,\ZeemanHanleHWHM)$
for the apparent spin lifetime $\tauSpinApp(x)=1/\ZeemanHanleHWHM$.
Since $\tauSpinApp$ does not depend on the prefactor $j_{x}^{z}$
in $s_{z}$, it is a function of $\tauSpinxyzFac,\,\brelaxNoDim,\,\tauSpinGeomMean,\, x,\, D$.
From dimensional analysis we see that \begin{equation}
\tauSpinApp=\tauSpinGeomMean\,\HWHMdimAnFunc_{\tauSpinxyzFac,\brelaxNoDim}\left(\frac{x}{\Ls}\right),\end{equation}
with some dimensionless function $\HWHMdimAnFunc_{\tauSpinxyzFac,\brelaxNoDim}$
that depends on the distance $\distSpinDiff=x/\Ls$ from the boundary
in units of the spin diffusion length.

\begin{figure}
\begin{center}\includegraphics[%
  width=70mm,
  keepaspectratio]{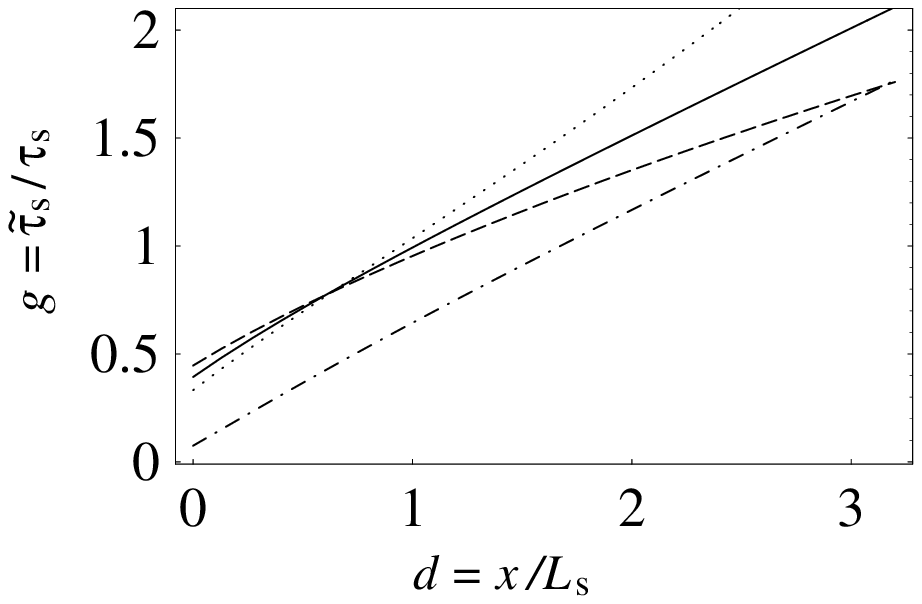}\vspace{-4mm}${}$\end{center}

\caption{\label{cap:g-plot}Apparent spin relaxation time $\tauSpinApp$ as
function of distance $x$ from boundary, plotted for $\brelaxNoDim=0$
and $\tauSpinxyzFac=1$ (solid line), $\tauSpinxyzFac=\sqrt{2}$ (dashed),
$\tauSpinxyzFac=1/\sqrt{2}$ (dotted) and for $\brelaxNoDim=2$ and
$\tauSpinxyzFac=1$ (dashed-dotted). While details depend on the microscopic
parameters (see text), generally the apparent spin lifetime is reduced
when considering the Hanle curves close to the boundary, even if the
spin relaxation rate is position-independent. }
\end{figure}

Using Eq.~(\ref{eq:szofxGen}), we evaluate $\HWHMdimAnFunc_{\tauSpinxyzFac,\brelaxNoDim}\left(\distSpinDiff\right)$
numerically and plot it in Fig.~\ref{cap:g-plot}. For example, $\HWHMdimAnFunc_{1,0}\approx0.43+0.52\distSpinDiff$
within $5\%$ and for $\distSpinDiff<10$. Most importantly, the {}``apparent''
spin relaxation time $\tauSpinApp$ shows a strong spatial dependence,
even if the underlying spin relaxation rate is homogeneous. In particular,
this means that $\tauSpinApp$ is roughly \emph{four times smaller}
near the boundary than several (three to four) spin diffusion lengths
away. This is in agreement with the experiments \cite{KatoSpinHall,ShiInhomogeneous,SternZnSe}
where a similar factor was observed \cite{footnoteStern}. 

Furthermore, note that at finite distances $x$, the Hanle curve can
develop {}``sidelobes,'' where $s_{z}$ changes sign, see Fig.~\ref{cap:hanle}.
This is a well-known effect and such sidelobes were detected electrically
in Johnson-Silsbee geometries for a fixed injector-detector spacing
$x$ \cite{JedemaSidelobe,ValenzuelaSH}. Additionally, in the regime
$\tauSpinInPlane\gg\tauSpinz$ and for a fixed $x$, the polarization
at finite fields can have a larger magnitude (but opposite sign) that
the main peak at zero fields, which can be understood as follows.
In the absence of spin precession (main peak), the spins will relax
rapidly with rate $\tauSpinz^{-1}$. However, the precessing spins
corresponding to the sidelobes relax with a lower average rate and
thus contribute with a larger signal, effectively filtering spins
that have precessed by an angle of $\pi$.

An important question is what happens at the boundary of a homogeneous
sample, namely if there are spin relaxations processes due to the
boundary. Such processes, on length scales shorter than $\Ls$, are
included here via $\invBDdecayLength$. By measuring the spatially
dependent Hanle curves and by fitting with Eq.~(\ref{eq:szofxGen})
(or by comparing with Fig.~\ref{cap:g-plot}), one can extract $\invBDdecayLength$
and therefore gain access to the relaxation at the boundary, even
if it occurs on a much shorter length scale than spatial resolution
of $s_{z}(x)$. Finally, for an inhomogeneous sample, a local probe
of the spin lifetime is desirable. While it is now clear that spin
diffusion can make such a measurement difficult in the steady state,
one could instead use a time-resolved (pump-probe) measurement to
determine $\tauSpinGeomMean(x)$. 

Instead of extracting the parameters of Eq.~(\ref{eq:Precession})
and (\ref{eq:Boundary}) by fitting to Eq.~(\ref{eq:szofxGen}),
one can find some parameters more directly as follows. First note
that for $B=0$, one can extract the decay length $\Lsz=\sqrt{D\tauSpinz}$
from $s_{z}(x)$. Next, the width of the Hanle curve contains information
about spin relaxation, and we access it via the curvature at the origin,
$\curvHanle(x)=(\partial^{2}s_{z}/\partial B^{2})/s_{z}|_{B=0}$.
(The normalization of $\curvHanle$ eliminates effects of a spatially
dependent detection sensitiviy on $s_{z}$.) Since the Hanle curve
becomes narrower when moving away from the boundary, the curvature
increases and from Eq.~(\ref{eq:szofxGen}), we find \begin{equation}
\left.\frac{1}{\curvHanle}\,\frac{\partial\curvHanle}{\partial x}\right|_{x=0}=\frac{1}{\Lsz}+\invBDdecayLength,\label{eq:curvToQs}\end{equation}
which does not explicitly depend on $\tauSpinxyzFac$ or $g^{*}$.
Because $\Lsz$ can be determined independently, Eq.~(\ref{eq:curvToQs})
provides a convenient way to access the spin relaxation $\invBDdecayLength$
at the boundary. 

In addition to the extrinsic spin-orbit interaction, leading to the
spin Hall effect considered above, there is also intrinsic spin-orbit
interaction that couples to the spin as an effective field $\vec{b}(\vec{k})$,
depending on the wave vector $\vec{k}$. In Eq.~(\ref{eq:Precession}),
we do not take this field into account explicitly; however, it does
contribute to the spin relaxation rate $\tauSpinGeomMean^{-1}$. Also,
this field can lead to additional spin polarization induced by the
electric field---for example, for a two-dimensional system with Rashba
spin-orbit interaction, this polarization is along the $x$ axis \cite{IvchenkoPikusSpinPol,VaskoPrima79,LNMSpinPolarization,Edelstein90,ALGPspinPol91,BernevigStrain,ERH_Polarzation};
however, it is not relevant in our discussion of $s_{z}$, since $s_{x}$
does not couple to $s_{z}$ or $s_{y}$ in Eq.~(\ref{eq:Precession}).
In a naive model, one can understand this polarization as arising
from the field $\driftField=\left\langle \vec{b}(\vec{k})\right\rangle $
averaged over all carriers, which drift in the electric field with
a finite $\left\langle \vec{k}\right\rangle $. For Rashba spin-orbit
interaction, $\driftField$ is in-plane and perpendicular to $\vec{E}$,
i.e., in our case $\driftField\parallel\vec{B}$.

In addition to the $s_{x}$ polarization, $\driftField$ contributes
as a spin precession term in the Bloch equation. Because it is parallel
to $\vec{B}$, its contribution can be absorbed into $\ZeemanHanle$
and it leads to a shifted Hanle curve $s_{z}(B)$ with the maximum
moved away from $B=0$. Experimentally, the expected shift of the
Hanle curve $s_{z}(B)$ was reported for strained three-dimensional
$n$-GaAs systems (where a spin-orbit coupling with the same form
as the Rashba term is present \cite{PikusTitkov}), while for unstrained
samples one sees that $\driftField=0$ due to the cubic symmetry and
there is no shift~\cite{KatoSpinHall}. Note that this naive model
can break down for more general transport mechanisms \cite{ERH_Polarzation},
which can lead to spin generation along $\driftField\times\vec{B}$,
but this expression vanishes in our configuration.

Furthermore, for Rashba spin-orbit interaction there are additional
precession terms around the $\vec{\hat{y}}$ axis that arise when
spins diffuse away from the edge~\cite{Burkov04,Mish04,AdagideliBauer,RashbaReview05}.
This would induce oscillations in $s_{z}(x)$ in addition to the one
shown in Fig.~\ref{cap:hanle}(d) and the combined effect can lead
to larger oscillation amplitudes. Since the precession length is on
the order of $\Ls$ in both cases, strictly speaking our model {[}Eq.~(\ref{eq:Precession}){]}
does not apply to a system with Rashba spin-orbit interaction---however,
no such $k$-linear intrinsic spin-orbit terms are present for a three-dimensional
system with cubic symmetry, which applies to the experiments of Ref.~\onlinecite{KatoSpinHall}
on unstrained samples. Finally, for two-dimensional systems, it was
argued that the Rashba spin-orbit interaction can change the magnitude
of extrinsic spin currents \cite{TseDasSarmaInEx,HankVignaleExtInt}
and would thus change the magnitude of the Hanle curves. For these
systems, also the importance of the intrinsic spin-orbit interaction
on the boundary conditions was studied \cite{BleibaumBC,TserkovnyakBoundary};
measuring the spatial dependence of the Hanle curves and using a property
analogous to Eq.~(\ref{eq:curvToQs}) can be used to test these predictions. 

In conclusion, we have found that in systems where spins are generated
at the boundary, the magnetic field dependence of the spin polarization
(Hanle curve) becomes spatially dependent even if the spin relaxation
rate $\tauSpin^{-1}$ is spatially homogenous. This leads to a reduction
of the {}``apparent'' spin lifetimes $\tauSpinApp$ near the edges
of a sample exhibiting the spin Hall effect, as was recently observed
experimentally \cite{KatoSpinHall}. We have provided an intuitive
picture for this effect: spins detected closer than $\Ls$ to the
edge were on average generated within a time less than $\tauSpin$
and relatively large magnetic fields would be required to suppress
them, corresponding to a small $\tauSpinApp$. Our description provides
a method for extracting the homogeneous spin relaxation rate and it
also allows to measure spin relaxation effects at the sample boundary.

We thank I. G. Finkler, B. I. Halperin, J. J. Krich, and E. I. Rashba
for many useful and insightful discussions. This work was supported
by NSF Grants No. DMR-05-41988 and No. PHY-01-17795.

\vspace{-5mm}${}$

\clearpage

\end{document}